\documentclass[letterpaper, 10 pt, conference]{ieeeconf}  

\IEEEoverridecommandlockouts                              
\usepackage{graphics} 
\usepackage{epsfig} 
\usepackage{mathptmx} 
\usepackage{times} 
\usepackage{amsmath} 
\usepackage{amssymb}  
\usepackage{tikz}
\usepackage{hyperref}
\usepackage[english]{babel}
\newtheorem{theorem}{Theorem}[section]
\newcommand{\ket}[1]{|#1\rangle}
\newcommand{\bra}[1]{\langle#1|}
\newcommand{\braket}[2]{\langle #1 | #2 \rangle} 

\title{\LARGE \bf
 Unitality Conditions on Subsystems in Quantum Dynamics

}

\author{Anumita Mukhopadhyay$^{*}$, Shibdas Roy and Arun Kumar Pati
\thanks{AM and SR are with Center for Quantum Engineering, Research and Education (CQuERE), TCG Centres for Research and Education in Science and Technology (TCG CREST), Kolkata 700091, India and Academy of Scientific and Innovative Research (AcSIR), Ghaziabad 201002, India. AKP is with Synergy Quantum India, Research and Innovation Park,
Indian Institute of Technology Delhi, New Delhi, India.
$^{*}$Email: {\tt\small anumitamukherjee455@gmail.com}}%
}

\begin{document}

\maketitle
\thispagestyle{empty}
\pagestyle{empty}

\begin{abstract}
It is known that non-unital noise such as the amplitude damping can sometimes increase quantum correlations, while unital noise such as the dephasing usually decreases quantum correlations. It is, therefore, important to delineate the conditions, when noise can enhance the quantumness of the system. Here, we show that if the noise acting on the system is unital (non-unital), then the noise acting on the environment must also be unital (non-unital), for the evolution to be unitary in the joint system-environment space. For example, if the first two qubits are treated as system and the third qubit is treated as environment, then both the system and the environment evolve unitally in case of a three-qubit GHZ state, and both of them evolve non-unitally in case of a three-qubit W state. Our result may be of interest in quantum information, and we anticipate it to be useful in various contexts, such as to better tackle noise in quantum computing and quantum information processing.
\end{abstract}

\section{Introduction}
Quantum systems are rarely isolated from external effects. Most of the time they interact with their surroundings. The environment is also a quantum system consisting of many degrees of freedom. A quantum system interacting with its environment is called an open quantum system. This interaction fundamentally changes how the system behaves. Environment can absorb energy, momentum, and information from the system \cite{information}, and it can also influence the system's evolution in significant ways. The joint quantum state of the combined system and environment evolves according to the standard rules of quantum mechanics, specifically, through a unitary transformation, which is a reversible operation. However, the system alone may be evolving non-unitarily. Whenever a quantum system evolves, its interaction with environment introduces noise in the system and then the evolution of the system takes place by noisy channel evolution \cite{davies,Schumacher}.  
Understanding open quantum systems is crucial because it reflects reality more accurately and explains phenomena like decoherence, where quantum properties are lost due to environmental interactions. 

Noise is generally considered to be detrimental to scaling up quantum computers, and avoiding noise in Noisy Intermediate Scale Quantum (NISQ) devices is challenging beyond a certain point. Thus, rather than avoiding noise, if we can manage to exploit it as a resource, then the challenges of NISQ-era quantum computation can be potentially circumvented. Existing literature already shows the possibility that noise can be used as a feature. Notable work in this field includes Ref.~\cite{cirac}, where dissipation is used as an alternative way of quantum computing and state engineering of large spectrum of highly correlated states without any coherent dynamics to complement it. The method of inducing quantum information capabilities using noise is also shown in Ref.~\cite{temporal}, which indicates that amplitude damping noise \cite{adc} can be utilized efficiently for Quantum Reservoir Computing (QRC) and information processing. Engineering dissipation for quantum information processing is also a new avenue, as established in Ref.~\cite{dissipaton}. Ref.~\cite{qcorr} shows non-unital noise channels \cite{non-unital} can create quantum correlations in multiqubit systems, while unital noise channels can introduce quantum correlations in multiqudit systems. Apart from such uses, noise can also be leveraged in quantum machine learning  \cite{qml,hqnet,bp}. Noise can be utilized to simulate open quantum systems effectively \cite{open,open2}. Noise can also be used as a resource for algorithmically solving open system dynamics on quantum computers \cite{qalgo}. Ref. \cite{req} illustrates that mixed entangled states can be arbitrarily more non-classical than separable and pure entangled states. Noise can also enhance entanglement in a robust manner in a quantum system \cite{noise}. 

Thus, we can see that noise can be potentially useful in many ways in quantum information. However, the precise conditions under which noise can enhance quantumness of the system is not too well understood. We think a systematic foundational approach from first principles to studying open quantum system dynamics is required. To this end, here we explore how the environment should evolve, for the system to evolve in a certain manner. In particular, the question we seek to answer is how the environment evolves, if the system evolves unitally. We show that for a general input state composed of a system and an environment, if the system evolves unitally, the environment must also evolve unitally, and when the system evolves non-unitally, then the environment must also evolve non-unitally. We illustrate our result for specific examples of the Bell state, the 3-qubit GHZ state \cite{ghz}, and the 3-qubit W state \cite{W}. We see that for the Bell state, created from $\ket{00}$, the system and the environment, being one qubit each, both evolve unitally. Similarly, for the GHZ state, created from $\ket{000}$, taking the first two qubits as system and the third one as environment, we see that the system and the environment both evolve unitally. In contrast to these two cases, taking the first two qubits as system and the third one as environment in the W state, created from $\ket{100}$, the system and the environment both evolve non-unitally. Moreover, using the metric named Relative Entropy of Quantumness (REQ) \cite{req1} to quantify the quantumness of a system, we note that the non-unital noise in the W state increases the quantumness of the system, while the non-unital noise acting on the environment does not increase its quantumness. It, therefore, remains to narrow down to further precise conditions for when non-unital noise enhances quantumness, as part of future work.

\section{Mathematical Background}
The evolution of open quantum systems 
can be modeled using the Kraus operator sum representation that describes the quantum operation which captures the interaction between the system and environment without explicitly needing to model the environment's state or the details of the interaction Hamiltonian. 
In our work, we assume the system-environment composite quantum system as a closed system evolving unitarily and thus delve deeper into understanding how the quantum channels or noise operators acting on the system and environment behave. We can take the input quantum state as $\rho_{S} \otimes \rho_{E}$ transforming by unitary $U$ and partially trace over the environment which gives the reduced state of the evolved system as: 
\begin{equation*}
  \varepsilon(\rho_{S}) = {\rm Tr}_{E}[U (\rho_{S}\otimes \rho_{E})U^\dagger]   
\end{equation*}
 Let $\ket{a_k}$ be the orthonormal basis for the environment. Thus the above equation becomes: 
 \begin{equation*}
     \varepsilon(\rho_{S}) = \sum_k \bra{a_k}U (\rho_{S}\otimes \ket{a_0}\bra{a_0})U^\dagger\ket{a_k} = \sum_k E_k \rho_{S}E_k ^\dagger,    
 \end{equation*}
where $E_k \equiv \bra{a_k}U\ket{a_0}$ is the operator on the Hilbert space of the system known as the Kraus Operators \cite{NC,kraus,choi,Stinspring}. They need to satisfy completeness relation given by $\sum_k E_k ^\dagger E_k = \mathbb{I}$. It is satisfied only when the quantum operation is trace-preserving, given as ${\rm Tr}(\varepsilon(\rho))=1$. Thus, Kraus operators are Completely Positive Trace Preserving (CPTP) maps \cite{pechukas,sudarshan,Brodutch,shabani,buscemi2}. Kraus operator representation is not a unique description. For a quantum operation we can have different Kraus representations related by unitary freedom \cite{preskill}. When the Kraus operators satisfy the relation $\sum_kE_kE_k ^\dagger = \mathbb{I}$ along with the completeness relation, we name that quantum channel as unital channel, and when $\sum_kE_k E_k ^\dagger \neq \mathbb{I}$ then the channel is called non-unital.

Unital channels are those, which leave the maximally mixed sates invariant, and are also known as doubly stochastic \cite{Buscemi_2006}. 
If $T: M_d \xrightarrow{} M_d$ is a linear map on $d \times d$ matrices, then CP enables Kraus decomposition:
\begin{equation*}
    T(\rho)= \sum_i A_i \rho A_i^\dagger,
\end{equation*}
where $\sum_i A_i ^\dagger A_i = I$. The channel is called unital, if $T(I)=I$, and including the trace preserving property, a unital channel is \textbf{doubly stochastic} CP map. Here, $M_d$ is a vector space equipped with $\langle A,B \rangle := {\rm Tr}[A^\dagger B]$, and forms a Hilbert space.
Then, the Kraus operators have the doubly stochastic form, $A_i:=\sqrt{p_i}U_i$, such that the map $T$ acts on $\rho$ as follows:
\begin{equation*}
    T(\rho)=\sum_i p_i U_i \rho U_i^\dagger,
\end{equation*}
 where $U_i U_i^\dagger = I$, $p_i > 0 \,\ \forall i$, and $\sum_i p_i = 1$.

A unital quantum channel preserves the average of the system states. On the other
hand, for a non-unital quantum channel, this is not the case. For example, in the case of a single qubit
the difference between unital and
non-unital channels is that the non-unital channels do not preserve the average
state in the center of the Bloch sphere.  One can check that the unital channels shrink the Bloch sphere in different
directions with the center preserved.  The non-unital quantum channel can not only shrink the
original Bloch sphere, but can move the center of the ball from the origin of the Bloch 
sphere. In the sequel, we present our main result.
 
 \section{Result}
Let $U$ be the unitary evolution acting on a joint system-environment composite system as in Fig \ref{fig:1}. Let the initial state be a general arbitrary quantum state $\rho_{SE}$, such that $\rho_S =\sum_j q_j \ket{\psi_j}\bra{\psi_j}$, and $\rho_E = \sum_j p_j \ket{a_j}\bra{a_j}$. Then, the noise acting on the system $S$ can be characterized by Kraus operators, $K_i = \sum_j\sqrt{p_j}\bra{a_i}U\ket{a_j}$, and that acting on environment $E$ can be characterized by Kraus operators, $L_j = \sum_k \sqrt{q_k}\bra{\psi_j}U\ket{\psi_k}$. For $K_i$ and $L_j$ to be valid Kraus representations, they need to satisfy completeness relations, i.e.~$\sum_iK_i^\dagger K_i = \mathbb{I}$ and $\sum_j L_j^\dagger L_j= \mathbb{I}$. Assuming the noise on the system to be unital, i.e.~$\sum_i K_i K_i^\dagger= \mathbb{I}$, we want to verify if the noise on the environment is also unital, i.e.~$\sum_j L_j L_j ^\dagger=\mathbb{I}$.

 \begin{theorem}
 If we assume the noise channel acting on the system to be unital, i.e.,~the Kraus operators of the noise channel on the system satisfy $\sum_iK_i K_i^\dagger= \mathbb{I}$, then the noise channel acting on the environment will also be unital, i.e.,~the Kraus operators of the noise channel on the environment will also satisfy $\sum_jL_j L_j ^\dagger=\mathbb{I}$, in order for the joint system-environment quantum state to evolve noiselessly via the unitary $U$.
\end{theorem}

\begin{figure}
     \centering
      \begin{tikzpicture}[scale=2.5]
        \draw[step=0.25cm,color=black] (0,0) rectangle (3,1)node[pos=.5] {Environment $E$ evolving unitally (non-unitally)} ;
      \draw (0,1) rectangle (3,2)node[pos=.5] {System $S$ evolving unitally (non-unitally)} ;
      \end{tikzpicture}
      \caption{A closed quantum system ($S+E$) evolving via joint unitary $U$.} \label{fig:1}
    \end{figure}
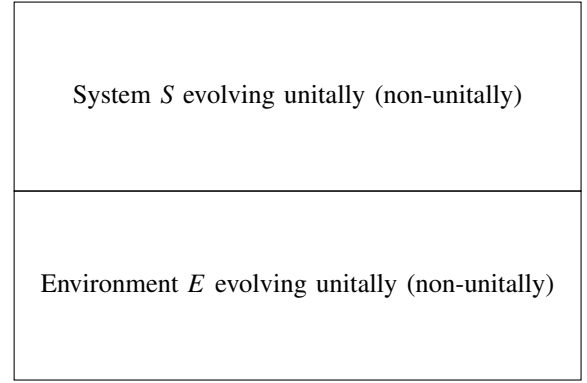
\begin{proof}
The system evolves as follows:
\begin{equation}
   \rho \xrightarrow{} T(\rho) = \sum_i p_i U_i \rho U_i^\dagger,
\end{equation}
where $\sum_ip_i =1$. 
The Kraus operators for the system and the environment, respectively, are:
\begin{equation}
\begin{split}
    K_i &= \sum_k \sqrt{p_k} \bra{a_i}U\ket{a_k}, \\
    L_j &= \sum_k \sqrt{q_k}\bra{\psi_j}U\ket{\psi_k}.
    \end{split}
\end{equation}
Since we assume the noise acting on the system to be unital, the unitary can be defined as $U=\sum_j U_j \otimes \Pi_j$, where $\Pi_j = \ket{a_j}\bra{a_j}$ are projectors for the environment $\rho_E$. Thus, we get: 
\begin{equation*}
\begin{split}
   K_i &= \sum_k \sqrt{p_k} \bra{a_i}\sum_j U_j \otimes \Pi_j \ket{a_k} \\
    &= \sum_{j,k}\sqrt{p_k} U_j\otimes\bra{a_i}\Pi_j\ket{a_k} \\
    &= \sum_{j,k}\sqrt{p_k} U_j\otimes\braket{a_i}{a_j}\braket{a_j}{a_k} \\
    &= \sum_j \sqrt{p_j} U_j \delta_{ij} \\ 
    &= \sqrt{p_i}U_i. 
\end{split}
\end{equation*}
For $K_i$ to be a valid CPTP noise, we must have:
\begin{equation}    \label{eq:unital_system}
\begin{split}
  \sum_i K_i ^\dagger K_i &= \sum_{i,j,k,l,m} \sqrt{p_k}\sqrt{p_m} \bra{a_k}U_j^\dagger\otimes \Pi_j\ket{a_i}\bra{a_i}U_l\otimes \Pi_l\ket{a_m} \\ \nonumber
   &= \sum_{i,j,k,l,m}\sqrt{p_k}\sqrt{p_m}U_j^\dagger U_l \bra{a_k}\Pi_j\ket{a_i}\bra{a_i}\Pi_l\ket{a_m} \\ \nonumber
    &= \sum_{i,j,k,l,m}\sqrt{p_k}\sqrt{p_m}U_j^\dagger U_l \braket{a_k}{a_j}\braket{a_j}{a_i}\braket{a_i}{a_l}\braket{a_l}{a_m}  \\ \nonumber
    &=\sum_{i,j,k,l,m}\sqrt{p_k}\sqrt{p_m} U_j^\dagger U_l \delta_{kj}\delta_{ji}\delta_{il}\delta_{lm}\\ \nonumber
    &= \sum_k p_k U_k^\dagger U_k \\ \nonumber
    &= \mathbb{I}.
\end{split}
\end{equation}
For the noise to be unital, we must have,
\begin{equation}
\begin{split}
   \sum_i K_i K_i ^\dagger &= \sum_{i,j,k,l,}\sqrt{p_k}\sqrt{p_m} \bra{a_i}U_j\otimes \Pi_j \ket{a_k}\bra{a_m}U_l^\dagger \otimes \Pi_l  \ket{a_i}\\
   &= \sum_{i,j,k,m}\sqrt{p_k}\sqrt{p_m}  U_j U_l^\dagger \bra{a_i}\Pi_j\ket{a_k}\bra{a_m} \Pi_l\ket{a_i} \\ 
    &= \sum_{i,j,k,m} \sqrt{p_k}\sqrt{p_m}U_j U_l^\dagger\braket{a_i}{a_j}\braket{a_j}{a_k}\braket{a_m}{a_l}\braket{a_l}{a_i} \\
    &= \sum_{k}p_k U_k U_k ^\dagger \\
    &= \mathbb{I}.
\end{split}
\end{equation}
Now, for $L_j$ to be a valid CPTP noise, we must have:
\begin{equation}\label{eq:unital_env}
\begin{split}
   \sum_j L_j ^\dagger L_j &= \sum_{i,j,k,l,n} \sqrt{q_k}\sqrt{q_n} \bra{\psi_k}U_i ^\dagger\ket{\psi_j}\bra{\psi_j}U_l\ket{\psi_n} \otimes \Pi_i \Pi_l \\
    &= \sum_{j,k,l,n} \sqrt{q_k}\sqrt{q_n}\bra{\psi_k}U_l ^\dagger\ket{\psi_j}\bra{\psi_j}U_l\ket{\psi_n} \Pi_l\\
    &= \mathbb{I},
\end{split}
\end{equation}
which is evident, since $\sum_j\ket{\psi_j}\bra{\psi_j}=\mathbb{I} $, $\braket{\psi_k}{\psi_n}= \delta_{kn}$, $U_l ^\dagger U_l=\mathbb{I}$, $\sum_l \Pi_l = \mathbb{I}$ and $\sum_k q_k=1$. The noise will be unital, if we have
\begin{equation}
\begin{split}
    \sum_j L_j L_j ^\dagger &= \sum_{i,j,k,l,n} \sqrt{q_k}\sqrt{q_n}\bra{\psi_j}U_i \ket{\psi_n}\bra{\psi_k}U_l^\dagger \ket{\psi_j} \otimes \Pi_i \Pi_l \\
    &= \sum_{j,k,l,n}\sqrt{q_k}\sqrt{q_n} \bra{\psi_j}U_l \ket{\psi_n}\bra{\psi_k}U_l^\dagger\ket{\psi_j} \Pi_l\\
    &= \mathbb{I},
\end{split}
\end{equation}
which evidently holds from (\ref{eq:unital_env}), since
\begin{equation}
\begin{split}
  \bra{\psi_j}U_l \ket{\psi_n}\bra{\psi_k}U_l^\dagger\ket{\psi_j}
  &= \bra{\psi_k}U_l ^\dagger\ket{\psi_j}\bra{\psi_j}U_l\ket{\psi_n},
\end{split}   
\end{equation}
where $\bra{\psi_j}U_l \ket{\psi_n}$ and $\bra{\psi_k}U_l^\dagger\ket{\psi_j}$ both are scalars.
Thus, the environmental Kraus operators evidently need to admit unitality, if the system Kraus operators admit unitality. In other words, if the noise acting on the system is unital (non-unital), then the noise acting on the environment must also be unital (non-unital).
\end{proof}

Our result can have interesting applications in quantum information. 
We know that doubly stochastic channels (which are unital and bistochastic) are often associated with entropy-increasing processes in quantum thermodynamics. For example, if a channel is unital then the von Neumann entropy of the system will always increase under the action of the channel. This means that if entropy of the system increases, then entropy of the environment will also increase. In the case of non-unital channel, the same cannot be said as the von Neumann entropy may increase or decrease.  Thus, the main result can be of use in characterization of reversible and irreversible processes in open system dynamics.
This can have applications in quantum error correction. Since unital noise does not introduce dissipation, it can often be corrected more easily than non-unital noise using quantum error correction (QEC) codes. 
This means that QEC codes can correct unital noise using standard syndrome measurement and recovery, whereas non-unital noise might require energy injection (e.g., using quantum control techniques) to restore lost quantum information.
This will be explored in more detail in the future.

\section{Examples}
\begin{itemize}
\item  Consider a $2$-qubit \textbf{Bell state} $\frac{1}{\sqrt{2}}[\ket{00}+ \ket{11}]$, created from $\ket{00}$, using the unitary:
\begin{equation*}
    \begin{split}
      U &= CNOT(H \otimes I)\\
      &=\frac{1}{\sqrt{2}}[\ket{00}\bra{00}+\ket{00}\bra{10}+\ket{01}\bra{01}+\ket{01}\bra{11}\\
      &+\ket{10}\bra{01}+\ket{11}\bra{00}-\ket{11}\bra{10}-\ket{10}\bra{11}]. 
    \end{split}
\end{equation*}
If we consider qubit $2$ as the system, the Kraus operators of the noise acting on the environment, i.e.~qubit $1$, are 
\begin{equation*}
    \begin{split}
        E_0 &=~_2\bra{0}U\ket{0}_2 \\
        &=\frac{1}{\sqrt{2}}[\ket{0}\bra{0}+ \ket{0}\bra{1}],
    \end{split}
\end{equation*}
and
\begin{equation*}
    \begin{split}
      E_1 &=~_2\bra{1}U\ket{0}_2\\
      &=\frac{1}{\sqrt{2}}[\ket{1}\bra{0}-\ket{1}\bra{1}],  
    \end{split}
\end{equation*}
and those of the noise acting on the system are 
\begin{equation*}
    \begin{split}
      S_0 &=~_1\bra{0}U\ket{0}_1\\
      &=\frac{1}{\sqrt{2}}[\ket{0}\bra{0}+\ket{1}\bra{1}],  
    \end{split}
\end{equation*}
and 
\begin{equation*}
    \begin{split}
       S_1 &=~_1\bra{1}U\ket{0}_1\\
       &=\frac{1}{\sqrt{2}}[\ket{0}\bra{1}+\ket{1}\bra{0}] .
    \end{split}
\end{equation*}
Clearly, we then have 
\begin{equation*}
    E_0E_0^\dagger + E_1E_1^\dagger = \ket{0}\bra{0}+\ket{1}\bra{1}= \mathbb{I},
\end{equation*}
and
\begin{equation*}
    S_0S_0^\dagger + S_1S_1^\dagger =\ket{0}\bra{0}+\ket{1}\bra{1}= \mathbb{I},
\end{equation*}
implying that {\it both the noises are unital.}

    \item Consider a $3$-qubit \textbf{GHZ state}$\frac{1}{\sqrt{2}}[\ket{000}+\ket{111}]$, created from $\ket{000}$, using the unitary:
    \begin{equation*}
        \begin{split}
          U&= (I \otimes CNOT)(CNOT \otimes I)(H \otimes I \otimes I )\\
          &= \frac{1}{\sqrt{2}}[\ket{000}\bra{000}+\ket{000}\bra{100}+\ket{001}\bra{001}+\ket{001}\bra{101}\\
          &+ \ket{010}\bra{011}+\ket{010}\bra{111}+\ket{011}\bra{010}+\ket{011}\bra{110}\\
          &+ \ket{100}\bra{010}-\ket{100}\bra{110}+\ket{101}\bra{011}-\ket{101}\bra{111}\\
          &+ \ket{110}\bra{001}-\ket{110}\bra{101}+\ket{111}\bra{000}-\ket{111}\bra{100}].  
        \end{split}
    \end{equation*}
 If we consider qubit $3$ as the environment, the Kraus operators of the noise acting on the system, i.e.~qubits $1$, $2$, are 
 \begin{equation*}
     \begin{split}
       S_0 &=~_3\bra{0}U\ket{0}_3\\
       &= \frac{1}{\sqrt{2}} 
    [\ket{00}\bra{00}+\ket{00}\bra{10}+\ket{10}\bra{01}-\ket{10}\bra{11}],  
     \end{split}
 \end{equation*}
and 
\begin{equation*}
    \begin{split}
       S_1 &=~_3\bra{1}U\ket{0}_3 \\
       &= \frac{1}{\sqrt{2}} 
    [\ket{01}\bra{01}+\ket{01}\bra{11}+\ket{11}\bra{00}-\ket{11}\bra{10}], 
    \end{split}
\end{equation*}
and those of the noise acting on the environment are
\begin{equation*}
    \begin{split}
      E_0 &=~ _{12}\bra{00}U\ket{00}_{12}\\ 
    &= \frac{1}{\sqrt{2}}[\ket{0}\bra{0}+ \ket{1}\bra{1}],  
    \end{split}
\end{equation*}
\begin{equation*}
    \begin{split}
        E_1 =~_{12}\bra{01}U\ket{00}_{12} = 0,
    \end{split}
\end{equation*}
\begin{equation*}
    \begin{split}
       E_2 =~ _{12}\bra{10}U\ket{00}_{12} = 0, 
    \end{split}
\end{equation*}
and
\begin{equation*}
    \begin{split}
        E_3&=~ _{12}\bra{11}U\ket{00}_{12} \\
        &= \frac{1}{\sqrt{2}}[\ket{0}\bra{1}+ \ket{1}\bra{0}].
    \end{split}
\end{equation*}
Clearly, we then have 
\begin{equation*}
    S_0S_0^\dagger + S_1S_1^\dagger =\ket{00}\bra{00}+\ket{01}\bra{01}+\ket{10}\bra{10}+\ket{11}\bra{11}=  \mathbb{I},
\end{equation*}
and
\begin{equation*}
    E_0E_0^\dagger + E_1E_1^\dagger + E_2E_2^\dagger + E_3E_3^\dagger =\ket{0}\bra{0}+\ket{1}\bra{1}=\mathbb{I},
\end{equation*}
implying that {\it both the noises are unital.}

  \item Consider a $3$-qubit \textbf{W state} $\frac{1}{\sqrt{3}}[\ket{001}+\ket{010}+\ket{100}]$, created from $\ket{100}$, using the unitary:
  \begin{equation*}
      \begin{split}
          U&= \ket{000}\bra{000}+
     \frac{1}{{\sqrt{3}}}\ket{001}\bra{001}-
     \frac{1}{{\sqrt{3}}}\ket{001}\bra{010}\\
     &+\frac{1}{{\sqrt{3}}}\ket{001}\bra{100}
     -\frac{1}{{\sqrt{3}}}\ket{010}\bra{001}+
     \frac{1}{{\sqrt{3}}}\ket{010}\bra{011}\\ 
     &+\frac{1}{{\sqrt{3}}}\ket{010}\bra{100}+
     \ket{011}\bra{101}+\frac{1}{{\sqrt{3}}}\ket{100}\bra{010}\\
     &-\frac{1}{{\sqrt{3}}}\ket{100}\bra{011}+\frac{1}{{\sqrt{3}}}\ket{100}\bra{100}+\ket{101}\bra{110}\\
     &+  \frac{1}{{\sqrt{6}}}\ket{110}\bra{001}+\frac{1}{{\sqrt{6}}}\ket{110}\bra{010}+\frac{1}{{\sqrt{6}}}\ket{110}\bra{011}\\
     &+\frac{1}{{\sqrt{2}}}\ket{110}\bra{111} 
     +\frac{1}{{\sqrt{6}}}\ket{111}\bra{001}+\frac{1}{{\sqrt{6}}}\ket{111}\bra{010}\\
     &+\frac{1}{{\sqrt{6}}}\ket{111}\bra{011}-\frac{1}{{\sqrt{2}}}\ket{111}\bra{111}.
      \end{split}
  \end{equation*}
 If we consider qubit $3$ as the environment, the Kraus operators of the noise acting on the system, i.e.~qubits $1$, $2$, are 
 \begin{equation*}
     \begin{split}
       S_0 &=~_3\bra{0}U\ket{0}_3\\ 
     &= \ket{00}\bra{00}+\frac{1}{{\sqrt{3}}}\ket{01}\bra{10}+\frac{1}{{\sqrt{3}}}\ket{10}\bra{01}+\frac{1}{{\sqrt{3}}}\ket{10}\bra{10}\\
     &+\frac{1}{{\sqrt{6}}}\ket{11}\bra{01},
     \end{split}
 \end{equation*}
 \begin{equation*}
     \begin{split}
       S_1 &=~ _3\bra{1}U\ket{0}_3 \\
       &=-\frac{1}{{\sqrt{3}}} \ket{00}\bra{01}+\frac{1}{{\sqrt{3}}} \ket{00}\bra{10}+ \ket{10}\bra{11}+ \frac{1}{{\sqrt{6}}} \ket{11}\bra{01},    
     \end{split}
 \end{equation*}
 and those of the noise acting on the environment are 
 \begin{equation*}
     \begin{split}
        E_0 &=~_{12}\bra{00}U\ket{10}_{12}\\ 
 &= \frac{1}{{\sqrt{3}}} \ket{1}\bra{0}, 
     \end{split}
 \end{equation*}
 \begin{equation*}
     \begin{split}
        E_1 &=~ _{12}\bra{01}U\ket{10}_{12} \\
 &= \frac{1}{{\sqrt{3}}} \ket{0}\bra{0} + \ket{1}\bra{1}, 
     \end{split}
 \end{equation*}
 \begin{equation*}
     \begin{split}
        E_2 &=~ _{12}\bra{10}U\ket{10}_{12}\\
        &= \frac{1}{{\sqrt{3}}} \ket{0}\bra{0}, 
     \end{split}
 \end{equation*}
and 
\begin{equation*}
    \begin{split}
        E_3 =~ _{12}\bra{11}U\ket{10}_{12} = 0.
    \end{split}
\end{equation*}
Clearly, we then have 
\begin{equation*}
\begin{split}
    S_0S_0^\dagger + S_1S_1^\dagger &=\frac{5}{3}\ket{00}\bra{00}-\frac{1}{\sqrt{18}}\ket{00}\bra{11}+\frac{1}{3}\ket{01}\bra{01}\\
    &+\frac{1}{3}\ket{01}\bra{10}+\frac{1}{3}\ket{10}\bra{01}+\frac{5}{3}\ket{10}\bra{10}\\
    &+\frac{1}{\sqrt{18}}\ket{10}\bra{11}-\frac{1}{\sqrt{18}}\ket{11}\bra{00}\\
    &+\frac{1}{\sqrt{18}}\ket{11}\bra{10}+\frac{1}{3}\ket{11}\bra{11} \neq \mathbb{I},
\end{split}
\end{equation*}
and
\begin{equation*}
    E_0E_0^\dagger + E_1E_1^\dagger + E_2E_2^\dagger + E_3E_3^\dagger=\frac{2}{3}\ket{0}\bra{0}+\frac{4}{3}\ket{1}\bra{1} \neq \mathbb{I},
\end{equation*}
implying that {\it both the noises are non-unital.}   
\end{itemize}

\section{Discussion}
We now discuss how noise changes the quantity of quantumness in a given system for the above examples, by means of the metric Relative Entropy of Quantumness (REQ). To verify when the noise is actually increasing quantumness of the system, we can use the metric known as Relative Entropy of quantumness (REQ) \cite{req}. With respect to an appropriate reference basis, REQ can be used to quantify how much coherence or quantumness is present in a quantum state. 
 It is given by
 \begin{equation}\label{eq:req}
     Q(\rho) = \min_{\rm classical \, \sigma} S(\rho || \sigma),
\end{equation}
     where $S(\rho || \sigma)=  {\rm Tr}(\rho \log_2 \rho - \rho \log_2 \sigma )$. Here, $\sigma$ are strictly classically correlated states used as reference state. For example, let us take $\rho = \ket{+}\bra{+}$. Then, we take $\sigma = \ket{+}\bra{+}$, since $\rho$ is diagonal in the $\ket{+},\ket{-}$ basis. Then, we have   $ S(\rho || \sigma) = 0$
which implies that the state $\rho$ is classical with respect to the chosen basis $\ket{+}, \ket{-}$. But if we choose the basis as the computational basis, then we will have $\sigma = \frac{1}{2} [\ket{0}\bra{0}+\ket{1}\bra{1}]$. Then we will get:
\begin{equation*}
\begin{split}
    {\rm Tr}(\rho \log_2 \sigma) &= -1, \\
    {\rm Tr}(\rho \log_2 \rho) &= 0, \\
    S(\rho || \sigma) &= 1, 
    \end{split}
\end{equation*}
which implies that the state $\rho$ is maximally quantum for the single qubit case, with respect to the chosen computational basis. 
Now, if we take $\rho = \frac{3}{4} \ket{0}\bra{0} + \frac{1}{4} \ket{1}\bra{1}$, then we need to take $\sigma=\rho$. That is, when $\rho$ is diagonal in the chosen basis, we need to take $\sigma$ same as $\rho$, so that the REQ of this state is zero, implying that the state is classical in the chosen basis. In other cases, i.e.~when $\rho$ is not a diagonal matrix in the chosen basis, $\sigma$ should be taken as the maximally mixed state in that basis. Thus, REQ is a basis dependent measure.

\subsubsection{GHZ state}
Given $\ket{GHZ}= \frac{1}{\sqrt{2}}[\ket{000}+\ket{111}]$, we take $\rho=\ket{GHZ}\bra{GHZ}$, and $\sigma$ as the maximally mixed state $\sigma= \frac{\mathbb{I}}{2^3}$ in the computational basis. Thus, from \eqref{eq:req}, we get
\begin{equation*}
    {\rm Tr}[\rho \log_2 \rho]=0, 
\end{equation*}
\begin{equation*}
    {\rm Tr}[\rho \log_2 \sigma] = -3, 
\end{equation*}
giving REQ of GHZ state as $3$. 
Now, if we trace out the environment, i.e.~third qubit, we get $\rho_{GHZ}^S=\frac{1}{2}[\ket{00}\bra{00}+\ket{11}\bra{11}]$. Computing REQ of $\rho_{GHZ}^S$ is done with respect to $\sigma=\frac{1}{2}[\ket{00}\bra{00}+\ket{11}\bra{11}] $, as $\rho_{GHZ}^S$ is a diagonal density matrix. Hence, $S(\rho_{GHZ}^S||\sigma)=0$. Given that the initial system state was $\ket{00}$, we see in this example that the unital noise channel is not increasing the REQ of the system.
\subsubsection{W state}
Given $\ket{W} = \frac{1}{\sqrt{3}}[\ket{001}+\ket{010}+\ket{100}]$, we take $\rho=\ket{W}\bra{W}$ and $\sigma$ again as the maximally mixed state in the computational basis. This gives 
\begin{equation*}
    {\rm Tr}[\rho \log_2 \rho]= 0, 
\end{equation*}
\begin{equation*}
  {\rm Tr}(\rho \log_2 \sigma) = -3,
\end{equation*}
giving REQ of W state as 3.

Now, tracing out the environment i.e.~third qubit, we get $\rho_W ^S= \frac{1}{3}[\ket{00}\bra{00}+\ket{01}\bra{01}+\ket{01}\bra{10}+\ket{10}\bra{01}+\ket{10}\bra{10}]$. REQ of $\rho_W ^S$ will be calculated with respect to the two-qubit maximally mixed state $\sigma=\frac{1}{4}[\ket{00}\bra{00}+\ket{01}\bra{01}+\ket{10}\bra{10}+\ket{11}\bra{11}]$.
Thus we get
\begin{equation*}
    {\rm Tr}(\rho_W ^S \log_2 \rho_W ^S )= -0.92,
\end{equation*}
and 
\begin{equation*}
\begin{split}
    {\rm Tr}(\rho_W ^S \log_2 \sigma )&= \frac{2}{3}[-\ket{00}\bra{00}-\ket{01}\bra{01}\\
    &-\ket{01}\bra
    10-\ket{10}\bra{01}-\ket{10}\bra{10}] \\
    &=-2,
    \end{split}
\end{equation*}
which will give $S(\rho_W^S||\sigma)= -0.92 + 2 = 1.08$. Given that the initial system state was $\ket{10}$, we see that non-unital noise acting on the system part of W-state is increasing quantumness in this example. 
On the other hand, tracing out the system, i.e.~first two qubits, from the density matrix of W-state gives $\rho_W ^{E}= \frac{2}{3}\ket{0}\bra{0}+\frac{1}{3}\ket{1}\bra{1}$. We compute the REQ of this also with respect to the reference state $\sigma=\frac{2}{3}\ket{0}\bra{0}+\frac{1}{3}\ket{1}\bra{1}$, as $\rho_W^E$ is a diagonal density matrix. Thus, we get $S(\rho_W^E || \sigma)=0$. This implies that although the noise channel acting on the environment of W-state is non-unital, it is not increasing quantumness, given that the initial environment state was $\ket{0}$. Therefore, non-unital noise channel can increase quantumness but not always. In our future work we intend to narrow down to further precise conditions when non-unital noise channel increases quantumness and when it does not.

\section{Conclusion}
In this work, we explored the unitality conditions on subsystems in quantum dynamics. We found that when the system evolves unitally (non-unitally), then the environment must also evolve unitally (non-unitally), for the joint system-environment evolution to be unitary. We illustrated our result by means of specific examples. Moreover, using an appropriate quantifier for quantumness, called the Relative Entropy of Quantumness (REQ), we found that non-unital noise can enhance the quantumness of a system, but not necessarily always. This leaves us with having to further narrow down to precise conditions, under which non-unital noise can enhance quantumness of a system, as part of future work. Our findings here are anticipated to have fundamental significance for noise-assisted quantum computation.

\bibliographystyle{IEEEtran}
\bibliography{unital}

\end{document}